\documentclass[11pt]{article}

\usepackage{graphicx,amsmath,amsfonts,amssymb,dcolumn,amsthm,slashed,bbm,xspace,pgffor,tikz,mathbbol,latexsym,enumerate,amsmath}
\usepackage{soul}
\usepackage[colorlinks,citecolor=red,urlcolor=cyan,linkcolor=blue]{hyperref}
\usepackage[utf8]{inputenc}
\usepackage[english]{babel}
\usepackage{lmodern}
\usepackage{microtype}
\usepackage{cite}
\usepackage{adjustbox}
\usepackage[normalem]{ulem}

\numberwithin{equation}{section}

\begin{document}

\title{\textbf{From the massive Schwinger model to unimodular Euclidean gravity with fermions}}

\author{\textbf{Thales F.~Bittencourt}\thanks{thalesfernandesbittencourt@id.uff.br},~ \textbf{Rodrigo F.~Sobreiro}\thanks{rodrigo\_sobreiro@id.uff.br}\\
\textit{{\small UFF - Universidade Federal Fluminense, Instituto de F\'isica,}}\\
\textit{{\small Av. Litorânea, s/n, 24210-346, Niter\'oi, RJ, Brasil.}}}

\date{}
\maketitle

\begin{abstract}
The massive Schwinger model (two-dimensional electrodynamics coupled to Dirac fermions) in Euclidean spacetime is mapped onto two-dimensional unimodular gravity in the first-order formalism, also including fermions. The idea of the map relies almost entirely on the isomorphism between the gauge groups $U(1)$ and $SO(2)$. However, the resulting fermion-gravity coupling differs from the conventional form, explicitly violating parity and time-reversal symmetries. Moreover, these fermions exhibit an unconventional transformation behavior under $SO(2)$ transformations. We show, however, that these properties are absent on the gravitational side of the mapping when starting from an axial massive Schwinger model. Several explicit examples of the mapping are considered, including the constant electric field and point-particle systems. For these examples, we explicitly derive the corresponding gravitational manifolds and compute the associated metrics. Furthermore, we analyze the consistency of the mapping at the quantum level using the path-integral formalism and Becchi-Rouet-Stora-Tyutin (BRST) techniques. Our results demonstrate that the proposed mapping remains consistent at the quantum level, establishing a formal BRST-compatible field correspondence between quantum electrodynamics and quantum gravity, at least at the $U(1) - SO(2)$ gauge sector, which does not include the diffeomorphic features of gravity.

\end{abstract}

\maketitle

\newpage

\tableofcontents

\newpage

\section{Introduction}\label{intro}

Electrodynamics in two dimensions (1+1) exhibits several intriguing features. In particular, the Schwinger model \cite{Schwinger1962}, which describes two-dimensional quantum electrodynamics with massless fermions, is an exactly solvable quantum system via bosonization and exhibits dynamical mass generation for the gauge field due to the chiral anomaly. Its massive version \cite{Coleman1976,ColemanJackiwSusskind1975} is related to a sine-Gordon theory, which remains classically integrable \cite{Bialynicki-Birula:1971akx}. Both massless and massive versions display confinement-related phenomena \cite{Lowenstein1971, Abdalla:1991vua}, making them a lower-dimensional laboratory for investigating aspects of QCD in a quite controlled and simplified way.

Additionally, two-dimensional models also have well-known applications in condensed matter physics. Examples include the description of interacting electrons in one-dimensional systems through Luttinger liquids \cite{Haldane1981}, the use of bosonization techniques in strongly correlated condensed matter systems \cite{Coleman1975}, effective low-dimensional descriptions of carbon nanotubes \cite{Egger1997,Kane1997}, and the dynamics of electrons in quantum wires \cite{Yacoby1996}.

On the other hand, two-dimensional gravity is also of significant relevance, particularly in the pursuit of a quantum theory of gravity. In two dimensions, quantum gravity models are known to be consistent \cite{Polyakov:1987zb,Gross:1989vs,Witten:1989ig,Chamseddine:1989dm,Fukuma:1990jw,Kazama:1992ex}, providing valuable insights that motivate the search for a consistent quantization of gravity in four dimensions (see \cite{Sadovski:2024uhg} and references therein). Furthermore, geometrodynamical descriptions of certain gauge systems have been proposed \cite{Achucarro:1987vz,Witten:1988hc,Obukhov:1998gx,Sobreiro:2007pn,Sobreiro:2011hb,Assimos:2013eua,Assimos:2019yln,Assimos:2021eok}, in which gravity (geometrodynamics) serves as an alternative framework for describing standard gauge models. From a practical perspective, two-dimensional gravity also finds applications in condensed matter systems. For example, the Jackiw-Teitelboim (JT) model \cite{Teitelboim:1983ux,Jackiw:1984je}, which describes pure gravity coupled to a scalar field, serves as a useful framework for investigating certain aspects of four-dimensional black hole physics \cite{Mertens:2022irh,Cadoni:1994uf,Nayak:2018qej,Moitra:2019bub,Nojiri:2000ja}. Moreover, through its connection with the low-energy dynamics of the Sachdev-Ye-Kitaev (SYK) model \cite{Sachdev:1992fk,Kitaev2015}, the JT model provides a bridge between graphene systems and four-dimensional black holes \cite{PhysRevLett.121.036403,Franz2018}.

In a previous study \cite{Sobreiro:2021leg}, it was demonstrated that pure two-dimensional electrodynamics in Euclidean spacetime can be mapped onto a geometrodynamical model. This mapping leverages the isomorphism between the groups $U(1)$ (the gauge group of electrodynamics) and $SO(2)$ (the gauge group of Euclidean gravity) to establish a connection between electrodynamics and gravity in two dimensions. It is important to note that, at the present stage, this gauge-gravity correspondence is formulated in Euclidean spacetime (the metric is assumed to have Euclidean signature, $\sim\text{diag}(+,+)$). Such a choice is the natural one because the mapping $U(1) \longrightarrow SO(2)$ is a straightforward isomorphism. It is worth mentioning that it could be possible to extend the map to Minkowski spacetime by adding an additional step involving a Wick rotation, allowing the homomorphism map $U(1) \longrightarrow SO(1,1)$ to be realized. Nevertheless, such a possibility is left for future investigation. In the present work, we restrict our analysis to Euclidean spacetimes. Additionally, this mapping has also proven useful for obtaining the Jackiw-Teitelboim gravity from static three-dimensional Maxwell–Chern–Simons electromagnetism in another study \cite{Bittencourt:2025ttv}. It could also help broaden the range of approaches for applying quantum field theory techniques to condensed matter systems, such as those discussed in \cite{10.21468/SciPostPhysLectNotes.62} and the references therein. Beyond these applications, another interesting perspective concerns confinement in two-dimensional electrodynamics. In particular, it is natural to ask whether the confinement properties of the Schwinger model could admit an alternative geometrical interpretation within the gravitational formulation. We leave this question for future investigation, but such a description might provide new insights into confinement mechanisms.

Motivated by these considerations, in this work we apply the mapping developed in \cite{Sobreiro:2021leg} to two-dimensional electrodynamics coupled to Dirac fermions, mostly known as the massive Schwinger model. As a result, we obtain an unconventional gravity theory in which the gravity-fermion coupling explicitly violates parity (P) and time-reversal (T) symmetries. Furthermore, while the local $SO(2)$ gauge symmetry remains the standard one for gravitational fields—such as the zweibein and the spin connection—it differs for the fermions. The gravity-fermion coupling deviates from the usual form, and the gauge transformations for the fermions are also modified. The violation of P and T symmetries becomes evident in the structure of the fermion-gravity interaction. Regarding gauge transformations, we identify a novel type of gauge theory that describes fermions coupled to gravity in a fundamentally different way. Nevertheless, we show that by changing the starting electrodynamical theory to an axial coupled one, we can achieve usual $SO(2)$ gauge transformations on the gravitational side. This indicates that our mapping connects a P-and-T violating theory to a P-and-T symmetric theory and vice versa. Although this gravity model cannot be directly derived from four-dimensional electrodynamics, it can still be consistently defined in four dimensions. Moreover, we demonstrate that the proposed correspondence between two-dimensional electrodynamics and gravity remains valid at the quantum level, at least in a formal BRST-compatible scenario.

In addition, as a supplementary result, we provide some examples of the mapping at the classical level. Specifically, we show how the mapping works for some simple electromagnetic systems in two dimensions, such as the constant electric field, the single charged particle, and the one-dimensional dipole. The corresponding gravitational manifolds are then scrutinized, and their metrics are explicitly obtained. Interestingly, these geometries are found to be unimodular. Unimodular gravity, in which the determinant of the metric is fixed, has been extensively investigated as an alternative formulation of general relativity and as a framework with potentially different implications for the cosmological constant problem \cite{Anderson:1971pn,Unruh:1989db,HENNEAUX1989195,CarballoRubio:2022aca}. In the present work, however, unimodularity is not imposed as an independent condition. Rather, we show that it emerges naturally from the consistency of the electromagnetic-gravitational mapping by requiring the correspondence between gauge-invariant observables on both sides of the duality. The classical examples considered below provide explicit realizations of this general feature and further illustrate the consistency and richness of the proposed mapping.

This work is organized as follows. In Section \ref{MAP} we provide the details of the mapping from two-dimensional electrodynamics to two-dimensional gravity at the classical level. In Section \ref{PROP} we discuss the exotic properties of the gravity model obtained here. The classical examples of the mapping are displayed in Section \ref{examples}. In Section \ref{Q} we develop the quantization of the model. Finally, in Section \ref{conc} we display our conclusions.

\section{Mapping Electrodynamics into gravity}\label{MAP}

In this section, we consider the mapping between the massive Schwinger model (two-dimensional electrodynamics coupled to Dirac fermions) into a gravity theory. The essentials of the mapping in pure electrodynamics can be found in \cite{Sobreiro:2021leg}.

\subsection{Electrodynamics}

We start from the 2-dimensional Maxwell-Dirac action in the form\footnote{Exterior products are implicit in the notation.} (spacetime is assumed to be a two-dimensional manifold with Euclidean signature, say $\mathbb{R}^2$ for simplicity)
\begin{equation}
   S_e=\int\;\left[\frac{1}{\mathrm{e}^2}\left(\kappa F+\Theta*F-\frac{1}{2}\Theta\ast\Theta\right)+\bar{\psi}\left(i\gamma\ast D-m\ast1\right)\psi\right]\;,\label{ed0}
\end{equation}
with $F=dA$ being the electromagnetic field strength 2-form while $A$ is the gauge field 1-form. The quantity $\kappa$ is a parameter of mass dimension 2, and $\mathrm{e}$ is the charge of the electron, carrying dimension 1. The gamma matrices are collected as a 1-form object through\footnote{See Appendix \ref{Ap1} for the conventions and other definitions of gamma matrices in 2 dimensions.} $\gamma=\gamma_idx^i$, with $x\in\mathbb{R}^2$ and  $i,j,k\ldots\in\{0,1\}$. The field $\psi$ is a Dirac fermion while $\bar{\psi}=\psi^\dagger\gamma^0$ is its conjugate. The exterior covariant derivative is defined by $D=d-A$\footnote{The \(1/e^2\) term is not usually present because the electric charge is commonly introduced explicitly in the definition of the covariant derivative, \(D=d-ieA\). However, with the convention adopted in this work, \(D=d-A\), the electric charge is absorbed into the gauge field. This allows the introduction of a \(1/e^2\) factor, leading to a more convenient definition of the canonical dimensions of the fields and of the mapping that will be presented. The insertion of this constant in the action only amounts to a rescaling of Maxwell’s equations by an overall constant factor. All of this is achieved through the field redefinition \(A \mapsto A/e\).}. The 2-form field $\Theta$ is an auxiliary field whose field equation is simply $\Theta=F$. Thus, the usage of this equation recovers the usual form of the Maxwell term $\sim F\ast F$, see \cite{Obukhov:1998gx,Sobreiro:2021leg}. The remaining term, $\sim\kappa F$ is recognized as the Chern-Pontryagin topological density which violates P and T symmetries. Finally, $\ast$ stands for the dual Hodge operator. For completeness, the quantum numbers of all fields and parameters appearing in action \eqref{ed0} are displayed in Table \ref{table1}. The ghost number is included in view of BRST symmetry discussion we will provide in Section \ref{Q} \cite{Becchi:1975nq,Tyutin:1975qk,Piguet:1995er}.
\begin{table}[ht]
\centering
\begin{tabular}{|c|c|c|c|c|c|}
	\hline 
Fields & $A$ & $\Theta$ & $\psi$ & $\mathrm{e}$ & $\kappa$ \\
	\hline 
Dimension & $1$ & $2$ & $1/2$ & $1$ & $2$ \\ 
Rank & $(1,0,0)$ & $(2,0,0)$ & $(0,1,0)$ & $(0,0,0)$ & $(0,0,0)$ \\
Statistics & $-$ & $+$ & $-$ & $+$ & $+$ \\
\hline 
\end{tabular}
\caption{Quantum numbers of the fields in the electromagnetic theory: Dimension stands for the canonical dimension of the field/parameter; The $\pm$ sign stands for bosonic/fermionic statistics of the field variables and parameters; The Rank characterizes the set of ranks of the field variables and parameters by means of $(\mathrm{form}\;\mathrm{rank},\mathrm{spinor}\;\mathrm{rank},\mathrm{ghost}\;\mathrm{number})$.}
\label{table1}
\end{table}

Action \eqref{ed0} is invariant under $U(1)$ gauge transformations of the form
\begin{eqnarray}
\delta A&=&d\alpha\;,\nonumber\\
\delta\psi&=&\alpha\psi\;,\nonumber\\
\delta\bar{\psi}&=&-\alpha\bar{\psi}\;,\nonumber\\
\delta\Theta&=&0\;,\label{gt1}
\end{eqnarray}
with $\alpha=-\alpha^\ast$ being an infinitesimal local parameter. Clearly, the field strength 2-form is gauge invariant, $\delta F=0$.

Following \cite{Sobreiro:2021leg}, before we proceed in the map from \eqref{ed0} to a gravity theory, we need to introduce some extra fields in order to account for the correct number of independent fields in both theories. The total number of independent fields in the action \eqref{ed0} is five $(A_0,A_1,\Theta_{01},\psi_0,\psi_1)$. On the other hand, two-dimensional gravity in the first-order formalism is constructed with the zweibein 1-form $e^a$ and the spin-connection 1-form $\omega^{ab}$ (and spinors) with $a,b,c\ldots,h\in\{0,1\}$ being the $SO(2)$ gauge indices. Therefore, two-dimensional gravity carries eight independent fields $(e^0_0,e^0_1,e^1_0,e^1_1,\omega^{01}_0,\omega^{01}_1,\psi_0,\psi_1)$. This discrepancy is eliminated by introducing a boundary term of the form
\begin{equation}
S_b=\int\;d\left(\tilde{\varphi}\varphi-\tilde{\zeta}\zeta\right)\;,\label{sb0}
\end{equation}
where the properties of the new fields are collected in Table \ref{table2}. This extra term allows us to map eleven independent fields (on the electrodynamics side), namely the five already counted plus two scalar fields, $\tilde{\varphi}$ and $\tilde{\zeta}$, and four components of the 1-form fields $\varphi\equiv(\varphi_0,\varphi_1)$ and $\zeta\equiv(\zeta_0,\zeta_1)$, into another eleven independent fields (in the gravity side) yet to be addressed. Moreover, being a boundary term, it does not affect the field equations in the bulk of the system. Furthermore, if the system possesses a non-trivial boundary, the contribution of these fields is merely algebraic, \emph{i.e.} non-dynamical. Thence, the full action to be considered before the mapping is
\begin{equation}
    S_{eb}=S_e+S_b\;.\label{ed1}
\end{equation}
\begin{table}[ht]
\centering
\begin{tabular}{|c|c|c|c|c|}
	\hline 
Fields & $\tilde{\varphi}$ & $\varphi$ & $\tilde{\zeta}$ & $\zeta$ \\
	\hline 
Dimension & $0$ & $1$ & $0$ & $1$ \\ 
Rank & $(0,0,0)$ & $(1,0,0)$ & $(0,0,0)$ & $(1,0,0)$ \\
Statistics & $+$ & $-$ & $+$ & $-$ \\
\hline 
\end{tabular}
\caption{Quantum numbers of the auxiliary fields in the electromagnetic theory.}
\label{table2}
\end{table}

It is worth mentioning that the number of extra fields could be smaller. However, besides the fact that we want a boundary term, it is also desirable that this term and the new fields belong to the trivial sector of Becchi-Rouet-Stora-Tyutin (BRST) cohomology (see Section \ref{Q})\footnote{The mapping in \cite{Sobreiro:2021leg} is slightly different because it deals with a greater number of extra fields, totalizing seventeen fields on each side. Here, we simplify the mapping to consider only eleven fields on each side. Nevertheless, the BRST properties remain basically the same.}.

\subsection{From electrodynamics to gravity}\label{EQUIV}

It turns out that the electromagnetic action \eqref{ed1} can be mapped into a gravity action. In essence, we will employ the map defined in \cite{Sobreiro:2021leg}, which is based on the isomorphism $U(1)\longmapsto SO(2)$. Such a map allows one to identify the fields in the electrodynamical side with gravity fields, inducing thus a map from the Euclidean spacetime to a generic deformed two-dimensional  spacetime $\mathrm{M}$, namely $\mathbb{R}^2\longmapsto\mathrm{M}$ (In what follows, $x\in\mathbb{R}^2$ and $X\in\mathrm{M}$). 

Starting with the gauge group itself, an element $u=\exp\alpha\in U(1)$ can be mapped into an element $s=\exp(\alpha^{ab}\epsilon_{ab}E)\in SO(2)$ with indices $a,b,c,\ldots,h$ being associated with the tangent space $\mathrm{T}(\mathrm{M})$. The quantity $E=-\epsilon$ is the $SO(2)$ generator and $\epsilon\equiv\epsilon_{ab}$ is the matrix representation of the Levi-Civita symbol.  The Abelian rotation group $SO(2)$ being the local isometry group of $\mathrm{M}$. Therefore,
\begin{equation}
\alpha\longmapsto\alpha^{ab}\epsilon_{ab}\;,\label{map0}
\end{equation}
with $\alpha^{ab}\in\mathbb{C}$. Moreover, we impose the map between the electromagnetic fields to be of the form\footnote{Notice that the factors in \eqref{map1} are different from those employed in \cite{Sobreiro:2021leg}. These factors are actually quite free to be chosen, as discussed in \cite{Sobreiro:2021leg}.}
\begin{eqnarray}
\Theta(x)&\longmapsto&\mu^2\epsilon_{ab}e^a(X)e^b(X)\;,\nonumber\\
A(x)&\longmapsto&\epsilon_{ab}\omega^{ab}(X)\;,\label{map1}
\end{eqnarray}
with $\mu$ being a mass parameter, $[\mu]=1$, which can be constructed out from $\kappa$ and $\mathrm{e}$. For simplicity, one can take $\mu=\mathrm{e}$. Clearly, the second map in \eqref{map1} is the natural mapping between a $U(1)$ gauge field and an $SO(2)$ gauge field, the latter recognized as the spin-connection 1-form $\omega^{ab}=\omega^{ab}_\mu dX^\mu$, with $\mu,\nu...\in\{0,1\}$. In the first map in \eqref{map1}, the field $e^a$ stands for a gravitational zweibein 1-form field, $e^a=e^a_\mu dX^\mu$.

For the fields in the auxiliary action \eqref{sb0} we have
\begin{eqnarray}
(\tilde{\varphi}(x),\tilde{\zeta}(x))\equiv\tilde{\Phi}^a(x)&\longmapsto&\tau^a(X)\;,\nonumber\\
(\varphi(x),\zeta(x))\equiv\Phi^a(x)&\longmapsto&\phi^a(X)\;,\label{map2}
\end{eqnarray}

We remark that the mapping \eqref{map2} is different from the original version defined in \cite{Sobreiro:2021leg}\footnote{The present version is more accurate since it takes care of the statistical properties of the extra fields dealt here: they are all bosonic fields.}. Besides the dimension of the fields, we also have that $\phi^a=\phi^a_{\phantom{a}b}e^b$ (which we assume is a Lorentz-algebra-valued field, contributing then with only one independent component field, since $\phi_{ab}=-\phi_{ba}$). Therefore, we can see that, besides the eight independent fields already counted for two-dimensional gravity in the previous section, there are another two scalar fields $\tau^a\equiv(\tau^0, \tau^1)$ and one component of the 1-form field $\phi^a=\phi^a_{\phantom{a}b}e^b\equiv(\phi^0_{\phantom{a}1})$ totalizing the eleven promised fields for the gravitational part that match the eleven fields from the electrodynamical part.

One can also infer that \cite{Alvarez:2011gd}
\begin{equation}
\gamma_idx^i\longmapsto\gamma_adX^a=\gamma_ae^a(X)\;.\label{map4}
\end{equation}
For the spinor, the map is quite simple. Spinors in $\mathbb{R}^2$ are directly mapped into spinors in $\mathrm{M}$,
\begin{equation}
\psi(x)\longmapsto\psi(X)\;.\label{map5}
\end{equation}
It is worth mentioning that the Jacobian of the full map is not trivial, being given by 
\begin{equation}
J\propto\mathrm{det}^{3/2}(\phi^a_{\phantom{a}b})e\;,\label{jac0}
\end{equation}
with $e=\det(e^a_\mu)$, as presented in Appendix \ref{Ap2}\footnote{The Jacobian of the present mapping is different from the one in \cite{Sobreiro:2021leg} because the number of extra fields is different.}. Such non-triviality affects quantization attempts; see reference \cite{Sobreiro:2021leg} and Section \ref{Q}.

The quantum numbers of all fields on the gravity side are listed in Table \ref{table3}.
\begin{table}[ht]
\centering
\begin{tabular}{|c|c|c|c|c|c|c|}
	\hline 
Fields & $e^a$ & $\omega^{ab}$ & $\psi$ & $\phi^a$ & $\tau^a$\\
	\hline 
Dimension & $0$ & $1$ & $1/2$ & $1$ & $0$\\ 
Rank & $(1,0,0)$ & $(1,0,0)$ & $(0,1,0)$ & $(1,0,0)$ & $(0,0,0)$ \\
Statistics & $-$ & $-$ & $-$ & $-$ & $+$\\
\hline 
\end{tabular}
\caption{Quantum numbers of the fields in the gravity theory.}
\label{table3}
\end{table}

After a straightforward computation, the substitution of \eqref{map1}, \eqref{map2}, and \eqref{map4} into \eqref{ed1} leads to a gravity action of the form
\begin{eqnarray}
S_G&=&\int\frac{1}{8\pi G}\epsilon_{ab}\left(R^{ab}-\frac{\Lambda^2}{2}e^ae^b\right)+\int\bar{\psi}\left(i\gamma_ae^a\ast\nabla-\frac{m}{2}\epsilon_{ab}e^ae^b\right)\psi+\int\;d\left(\phi^a\tau_a\right)\;,\label{grav2}
\end{eqnarray}
with $G$ and $\Lambda$ being fixed by the constants $\mathrm{e}$ and $\kappa$ via
\begin{eqnarray}
 G&=&\frac{1}{8\pi}\frac{\mathrm{e}^2}{\left(\kappa+2\mathrm{e}^2\right)}\;,\nonumber\\
 \Lambda^2&=&\frac{2\mathrm{e}^4}{\left(\kappa+2\mathrm{e}^2\right)}\;,\label{param0}
\end{eqnarray}
and $R^{ab}=d\omega^{ab}$ is the curvature 2-form. The dimensions of the gravity parameters are given by $[G]=0$ and $[\Lambda]=1$. The covariant derivative $\nabla$ induced by the map is defined through
\begin{equation}
    \nabla=d-\epsilon_{ab}\omega^{ab}\;.\label{covdev0}
\end{equation}

Some direct properties of the gravity action \eqref{grav2} are discussed in the next section. 

It is worth mentioning that the mappings \eqref{map1} and \eqref{map4} are not merely a change of variables. They state that the fields affect spacetime directly. In other words, the map $U(1)\longmapsto SO(2)$ states that the electromagnetic gauge symmetry is mapped to the local spacetime isometries, allowing for a geometrical reinterpretation in terms of a dynamical coframe and an independent spin-connection. We emphasize that the first relation in \eqref{map1} fixes the volume two-form associated with the zweibein, rather than uniquely determining all components of \(e^a{}_\mu\). Accordingly, an electromagnetic configuration determines a class of gravitational configurations. The construction of a particular metric requires additional geometrical input, such as symmetry, topology, regularity, and, in the examples considered below, a torsion-free representative. The map \eqref{map4} is therefore an additional defining element of the correspondence: it promotes the fixed Cartesian coframe of the original Euclidean background to a soldering form on the target manifold.

\section{Parity violating unimodular gravity}\label{PROP}

It turns out that the gravity theory defined by \eqref{grav2} possesses some quite exotic properties, which we discuss in detail.

\subsection{The action}\label{sectiontheaction}

Starting with the pure gravity sector in \eqref{grav2}, namely $\sim\int(R-\Lambda)$, we immediately recognize the first term as the Einstein-Hilbert action which, in 2 dimensions, coincides with the Gauss-Bonnet topological term. Therefore, this term does not contribute to the field equations. The second term is clearly the cosmological constant term. Hence, the gravity sector itself essentially defines the Plateau problem. Still in the pure gravitation piece, the two-dimensional Newton and cosmological constants are fixed by the electrical charge and the Chern-Pontryagin constant defined in the original electromagnetic action. Remarkably, if $\kappa=0$ (no topological term in \eqref{ed0}), we get
\begin{eqnarray}
 G&=&\frac{1}{16\pi}\;,\nonumber\\
\Lambda^2&=&\mathrm{e}^2\;.\label{param1}
\end{eqnarray}
Thus, in that case, Newton's constant is fixed independently from any parameter, which is consistent with the fact that $G$ is dimensionless in 2 dimensions. Furthermore, a fixed parameter such as $G$ suggests that, at the quantum level, $G$ should not run. Since the zweibein and the spin connection are independent variables, and since the two-dimensional Einstein-Hilbert term is topological, variation with respect to \(\omega^{ab}\) does not yield the torsion-free condition \(T^a=0\). In the presence of fermions, this variation instead produces a constraint involving the fermionic current. Therefore, the mapped theory is naturally formulated in a first-order Riemann-Cartan geometry. A torsion-free Levi-Civita connection may be selected only as an additional restriction and is not a consequence of the mapped action.

The last term is a total derivative. If spacetime has no boundary (or if it lies in the infinity), this term does not contribute to the field equations whatsoever. If the spacetime boundary is non-trivial, then this term could be relevant at the boundary. However, it turns out that the contribution of this term at the boundary is purely algebraic. In fact, the same reasoning works for the Einstein-Hilbert term.

The fermionic term is certainly the most interesting one because it is not simply the usual Dirac action in curved spacetime. In fact, the interaction term reads (see relations \eqref{gamma4})
\begin{equation}
\mathcal{L}_I=e^a_\mu\epsilon_{bc}\omega^{bc}_\mu\bar{\psi}\gamma_a\psi=-e^a_\mu\omega^{bc}_\mu\bar{\psi}\gamma_a\gamma^3\sigma_{bc}\psi\;,\label{int0}
\end{equation}
with $\sigma_{bc}$ being the generators of the spinorial representation of $\mathfrak{so}(2)$ algebra, which clearly violates P and T symmetries\footnote{Under parity, $\mathsf{P}^0$ impose $x^0\to x^0$ and $x^1\to-x^1$, the zweibein and spin connection transform as $e^a_\mu\to\Lambda^a_{\phantom{a}b} e^b_\nu(\Lambda^{-1})^\nu_{\phantom{\nu}\mu}$ and $\omega^{ab}_\mu\to\Lambda^a_{\phantom{a}c}\Lambda^b{}d\omega^{cd}_\nu(\Lambda^{-1})^\nu_{\phantom{\nu}\mu}$, respectively, with $\Lambda^a_{\phantom{a}b}=\mathrm{diag}(1,-1)$. Consequently, the combination $e^a_\mu\epsilon_{bc}\omega^{bc}_\mu$ transforms as a pseudovector, while the fermionic vector current $\bar{\psi}\gamma_a\psi$ transforms as an ordinary vector. Their contraction therefore changes sign, $\mathcal{L}_I\xrightarrow{\mathsf{P}}-\mathcal{L}_I$, showing explicitly that the interaction term is not parity invariant. An analogous argument applies to time reversal, under which $\mathcal{L}_I$ also changes sign.}. Therefore, action \eqref{grav2} describes a parity violating gravity. It is true that the original electrodynamical action \eqref{ed0} already violates P and T symmetries due to the Chern-Pontryagin term. However, the Einstein-Hilbert action does not. The interacting term \eqref{int0} emerges even though we set $\kappa=0$. Thus, the parity violating term \eqref{int0} fundamentally originates from the mapping \eqref{map1}, specifically from the presence of the Levi-Civita density.

\subsection{Gauge symmetry}\label{G}

Another non-trivial point of the geometrodynamical sector is its gauge symmetry. Employing the maps discussed in Sect.~\ref{EQUIV} to the gauge transformations \eqref{gt1}, one readily finds
\begin{eqnarray}
\delta\omega^a_{\phantom{a}b}&=&d\alpha^a_{\phantom{a}b}\;,\nonumber\\
\delta e^a&=&\alpha^a_{\phantom{a}b}e^b\;,\nonumber\\
\delta\psi&=&\alpha^{ab}\epsilon_{ab}\psi\;,\nonumber\\
\delta\bar{\psi}&=&-\bar{\psi}\alpha^{ab}\epsilon_{ab}\;.\label{gt2}
\end{eqnarray}
Transformations of the zweibein and spin-connection are the usual  $SO(2)$ transformations; the gravitational fields transform normally under local $SO(2)$ rotations. However, fermions do not transform as expected under the very same rotations (see \eqref{gamma4}),
\begin{eqnarray}
\delta\psi&=&-\alpha^{ab}\gamma^3\sigma_{ab}\psi\;,\nonumber\\
\delta\bar{\psi}&=&\bar{\psi}\alpha^{ab}\gamma^3\sigma_{ab}\;.\label{gt3}
\end{eqnarray}
Therefore, the fermions in the mapped theory are not the typical fermions of the original one. Somehow, their nature changes due to the map. Transformations \eqref{gt3} resemble a local chiral transformation, but they are not. It is an $SO(2)$ transformation endowed with an extra $\gamma^3$ matrix. In fact, action \eqref{grav2} is not invariant under the usual global chiral transformations,
\begin{eqnarray}
\delta_c\psi&=&\gamma^3\theta\psi\;,\nonumber\\
\delta_c\bar{\psi}&=&-\bar{\psi}\gamma^3\theta\;,\label{ct1}
\end{eqnarray}
with $\theta$ being a global parameter\footnote{It is tempting to assume $\alpha^{ab}\sigma_{ab}$ as a scalar parameter just like $\theta$ and then associate transformations \eqref{gt3} to chiral ones. However, it is worth noting that each element in $\sigma_{ab}$ is a matrix as presented in Appendix \ref{Ap1}, so that could not be the case.}. Nevertheless, one can easily check that the action \eqref{grav2} remains invariant under transformations \eqref{gt2}. We recall that the origin of the mapping $U(1)\longmapsto SO(2)$ alone does not ensure the target theory to be a gravitational one; it is the existence of the zweibein. The zweibein is the field responsible for the identification of the gauge group $SO(2)$ with local isometries. Therefore, transformations \eqref{gt2} indeed are local $SO(2)$ transformations. Thence, we conclude that the kind of fermions we are dealing with here are quite different from ordinary.

We can take another path in order to comprehend the symmetry. To understand what is happening a bit more, let us take a look in transformations \eqref{gt3} when acting on the chiral projections of the spinor. We set the chiral projectors to be the Hermitian operator $P$ and its transpose\footnote{The $i$ factor is necessary because $(\gamma^3)^2=-1$, a consequence of our choice in working in Euclidean space.}
\begin{eqnarray}
    P&=&\frac{1}{2}\left(1+i\gamma^3\right)\;,\nonumber\\
    P^T&=&\frac{1}{2}\left(1-i\gamma^3\right)\;.
\end{eqnarray}
Thus, defining right and left handed components by
\begin{eqnarray}
\psi_R&=&P\psi\;,\nonumber\\
    \psi_L&=&P^T\psi\;,
\end{eqnarray}
one straightforwardly gets
\begin{eqnarray}
    \delta\psi_R&=&i\alpha^{ab}\sigma_{ab}\psi_R\;,\nonumber\\
    \delta\psi_L&=&-i\alpha^{ab}\sigma_{ab}\psi_L\;,\label{gt4}
\end{eqnarray}
which corresponds to an $SO(2)$ transformation of a spinor. Therefore, transformations \eqref{gt4} say that right-handed components transform normally with respect to $SO(2)$ transformations, but the left-handed components transform with the inverse. A wrong sign in a rotation may be seen as a reflection of the rotating axis, characterizing thus the effect of the parity violation of the model. 

Let us take a look at this effect in terms of $SO(2)$ rotations by considering the $0th$ coordinate as time. Suppose we have two particles with different hands, namely $p_R$ and $p_L$, both at rest with respect to observer $O_1$. Observer $O_2$ travels with respect to $O_1$ with speed $V$ to the right. The $SO(2)$ rotations for the speeds $v_R$ and $v_L$ (as observed by $O_2$) are simply (because they are at rest with respect to $O_1$)
\begin{equation}
    v_{R,L}=\mp V\;.
\end{equation}
Thus, instead of observing both particles going to the left with speed $-V$, observer $O_2$ sees the right-handed particle going to the left with speed $-V$ but the left-handed particle traveling to the right with velocity $V$. Such a bizarre effect can be associated with the parity violation of the model, meaning that besides an $SO(2)$ rotation, left-handed fermions also invert the axis of the coordinate system.

\subsection{From the axial massive Schwinger model}

So far, as discussed in the previous sections, we associated a P-and-T-symmetric theory with a P-and-T-violating theory because of the kind of coupling in the gravitational covariant derivative induced by the mapping. We could consider another type of electrodynamical theory to start from in order to achieve the opposite relation.

Let us restart with an axial massive Schwinger model, given by the following action
\begin{equation}
   S_{a}=\int\;\left[\frac{1}{\mathrm{e}^2}\left(\kappa F+\Theta*F-\frac{1}{2}\Theta\ast\Theta\right)+\bar{\psi}\left(i\gamma\ast \mathcal{D}-m\ast1\right)\psi\right]+ \int\;d\left(\tilde{\varphi}\varphi-\tilde{\zeta}\zeta\right)\;,\label{ed0axial}
\end{equation}
which is basically the same as \eqref{ed0} but with another covariant derivative: $\mathcal{D}=d-\gamma^3A$.

It is important to emphasize that, although there is now a global chiral gauge symmetry, the local $U(1)$ gauge symmetry remains intact. Hence, one may consistently apply the proposed mapping and obtain a gravitational action of the form
\begin{eqnarray}
S_G&=&\int\frac{1}{8\pi G}\epsilon_{ab}\left(R^{ab}-\frac{\Lambda^2}{2}e^ae^b\right)+\int\bar{\psi}\left(i\gamma_ae^a\ast\Delta-\frac{m}{2}\epsilon_{ab}e^ae^b\right)\psi+\int\;d\left(\phi^a\tau_a\right)\;,\label{grav2.0}
\end{eqnarray}
which is of the same form as \eqref{grav2} but with another type of covariant derivative: $\Delta=d-\sigma_{ab}\omega^{ab}$. This action has the following gauge transformations:
\begin{eqnarray}
\delta\omega^a_{\phantom{a}b}&=&d\alpha^a_{\phantom{a}b}\;,\nonumber\\
\delta e^a&=&\alpha^a_{\phantom{a}b}e^b\;,\nonumber\\
\delta\psi&=&\alpha^{ab}\sigma_{ab}\psi\;,\nonumber\\
\delta\bar{\psi}&=&-\bar{\psi}\alpha^{ab}\sigma_{ab}\;.\label{gt2.0}
\end{eqnarray}
which are clearly the proper $SO(2)$ gauge transformations. Therefore, with this map, we can achieve a P-and-T-symmetric theory from a P-and-T-violating theory as well.

\subsection{Unimodularity condition}

\begin{figure}[h]
    \centering
    \includegraphics[width=0.5\linewidth]{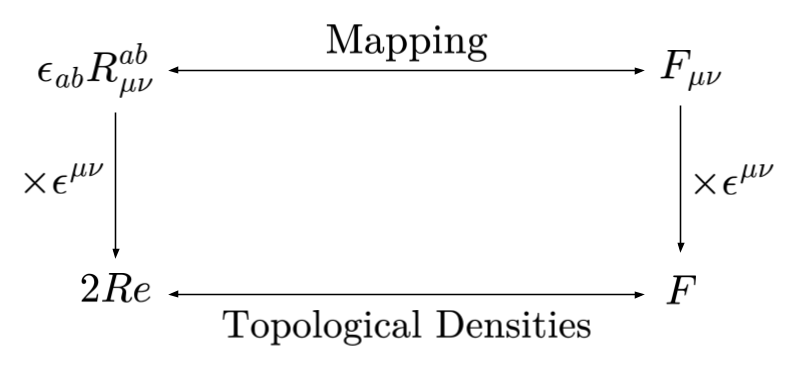}
    \caption{Topological densities identification.}
    \label{Pontryagin}
\end{figure}

Remarkably, it can be shown that the gravity found must be unimodular, \emph{i.e.}, $e=\det e^a_\mu=1$. With this purpose, we recall that in dual models, it is natural and desirable, to impose that observables in each theory are mapped into each other. Due to the second mapping in \eqref{map1}, it is straightforward to link the curvature to the electromagnetic tensor by means of\footnote{For ease of computation, tensor notation will be employed.}
\begin{equation}
\epsilon_{ab}R^{ab}_{\mu\nu}\longleftrightarrow F_{\mu\nu}\;.\label{fstrengths01}
\end{equation}
In order to achieve gauge invariant terms (typically related to possible observables), it is necessary to contract these structures with the proper objects in each theory: the corresponding Levi-Civita density works well for that. This is represented schematically in Figure \ref{Pontryagin}. Therefore, the curvature scalar is naturally identified with the Pontryagin density (both related to topological invariants on each side of the mapping).

For the sake of simplicity, we adopt the mathematical liberty to consider the relation \eqref{fstrengths01} an equality, just as it was done in \cite{Obukhov:1998gx}. Conventionalizing that $F_{01}=E$, with $E$ being the electric field, and from \eqref{fstrengths01}, one can immediately infer that
\begin{equation}
    R^{01}_{01}=\frac{1}{2}E\;,
\end{equation}
from which the curvature invariant can easily be derived by means of
\begin{equation}
    R=R^{ab}_{\mu\nu}e^\mu_a e^\nu_b=Ee\;.\label{invcurv01}
\end{equation}

On the other hand, following the idea of Fig.~\ref{Pontryagin}, we have
\begin{equation}
\epsilon_{ab}\epsilon^{\mu\nu}R^{ab}_{\mu\nu}=F_{\mu\nu}\epsilon^{\mu\nu}\;, \label{invs01}
\end{equation}
that yields\footnote{We have used the identity:
\begin{equation}
    \epsilon_{ab}\epsilon^{\mu\nu}=e\left(e^\mu_ae^\nu_b-e^\mu_be^\nu_a\right)\;.
\end{equation}} the relation
\begin{equation}
    R=\frac{E}{e}\;.\label{invcurv02}
\end{equation}
Comparing relations \eqref{invcurv01} and \eqref{invcurv02}, it is clear that we must demand
\begin{equation}
    \det e^a_\mu =1\;.\label{uni01}
\end{equation}
Thus,
\begin{equation}
    R=E\,.\label{RE01}
\end{equation}
Therefore, the gravity that is dual to two-dimensional electromagnetism must be unimodular. Hence, the diffeomorphism freedom of the model is restricted to the set of transverse diffeomorphisms. Moreover, since $R$ is an invariant of the theory, $E$ is also invariant in the unimodular formalism.

Let us comment a bit further on this: The identification \eqref{invs01}, connecting two different manifolds, is, in fact, responsible for the requirement \eqref{uni01}. Because we are considering the electromagnetic theory in Euclidean space and assuming Cartesian coordinates, the right hand side in \eqref{invs01} actually carries a unit determinant $\det\delta^a_\mu=1$. Therefore, for the mapping to be consistent, unimodular gravity must be required. 

Recalling that the electrodynamical theory is constructed in Euclidean spacetime for simplicity, one could ask what would happen if it were constructed over a curved spacetime (or even if we had used a non-Cartesian coordinate system). In this case, the determinant of such space, namely $\tilde{e}$, would appear on the right hand side of \eqref{invs01}. And from the gravity side, $\tilde{e}$ is a fixed function. The constraint on the determinant on the gravity side would read 
\begin{equation}
    e=\tilde{e}\;,\label{uni02}
\end{equation}
instead of \eqref{uni01}.
Nevertheless, the general condition for unimodularity is actually \eqref{uni02}. The case \eqref{uni01} is just the simplest case, see \cite{CarballoRubio:2022aca}.

\section{Simple examples: explicit mappings}\label{examples}

In this section, we provide some examples of mappings and consistency checks of the ideas of the mapping. For the sake of simplicity, we will not consider the spinor field, but we provide examples in terms of simple classical electric systems: The constant electric field; the static point charge; the electric dipole; and the point charge in uniform linear motion. In the following examples, for simplicity, we select torsion-free metric representatives in order to illustrate the curvature-field correspondence. This is an additional geometrical choice and is not imposed by the general first-order action.

\subsection{Constant electric field}

\begin{figure}[h]
    \centering
    \includegraphics[width=0.3\linewidth]{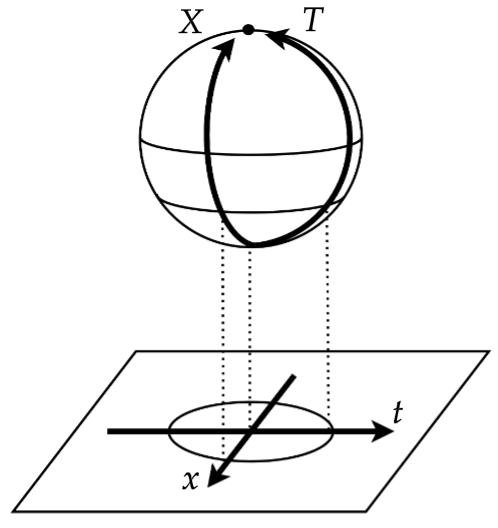}
    \caption{Sphere with its north pole extracted, $S^2\backslash\{N\}$, and the mediridians corresponding to radial lines emerging from the origin $(x=0,t=0)$ and the equator are depicted in the figure. The north pole is indicated by the black dot.}
    \label{spacetime1}
\end{figure}

Let us first explore the simple case of a constant electric field existing in the whole $\mathbb{R}^2$. From the previous section, specifically, relation \eqref{RE01} implies a 2-dimensional space of constant curvature ($E$ is constant), \emph{i.e.}, a sphere with its north pole extracted\footnote{The north pole must be rejected as part of the manifold because a full sphere is compact while the infinite plane is not.}. The formal mapping may be written as $\mathbb{R}^2\longmapsto S^2\backslash \{N\}$. Thus, a constant electric field in flat space can be represented by a sphere of radius given by 
\begin{equation}
    r=\sqrt{\frac{2}{E}}\;,\label{radius}
\end{equation} 
meaning that the stronger the electric field, the smaller the sphere is. The map is essentially realized by identifying all points at infinity with the north pole of the sphere. The usual map that allows such a realization is the inverse stereographic projection, but one can imagine a simple map realized by just wrapping the sphere with the plane. The mapped axes $X$ and $T$ corresponding to the axes $x$ and $t$, respectively, are then orthogonal meridians. The meridians correspond to radial lines originating from the point $(t=0,x=0)$ in the plane $\mathbb{R}^2$. In general, an arbitrary meridian will be a mixture of the $(t,x)$ coordinates. See Figure \ref{spacetime1}.

Additionally, we are restricted to unimodular metrics. The stereographic coordinates do not satisfy this condition. To achieve a unimodular metric, we can start from the usual spherical coordinate system in terms of the polar and azimuthal angles, namely, $\theta$ and $\phi$, respectively\footnote{We remark that the apparent horizon simply characterizes the fact that this coordinate system is ill defined at the south pole of the sphere.}. Thence, the usual angular coordinate system must be replaced by an equivalent metric whose determinant is precisely one. This change of coordinates is realized through
\begin{eqnarray}
    u & = & r^2\left(1-\cos\theta\right)\;,\\
    v & = & \phi\;.
\end{eqnarray}
Providing the unimodular metric,
\begin{equation}
    g_{\mu\nu}\equiv\begin{pmatrix}
f(u)^{-1} & 0 \\
0 & f(u)
\end{pmatrix}\;,\label{gsphere1}
\end{equation}
with
\begin{equation}
    f(u)=r^2\left[1-\left(1-\frac{u}{r^2}\right)^2\right]\;.
\end{equation}
The metric \eqref{gsphere1} immediately gives $e=1$.

\begin{figure}[h]
    \centering
    \includegraphics[width=0.15\linewidth]{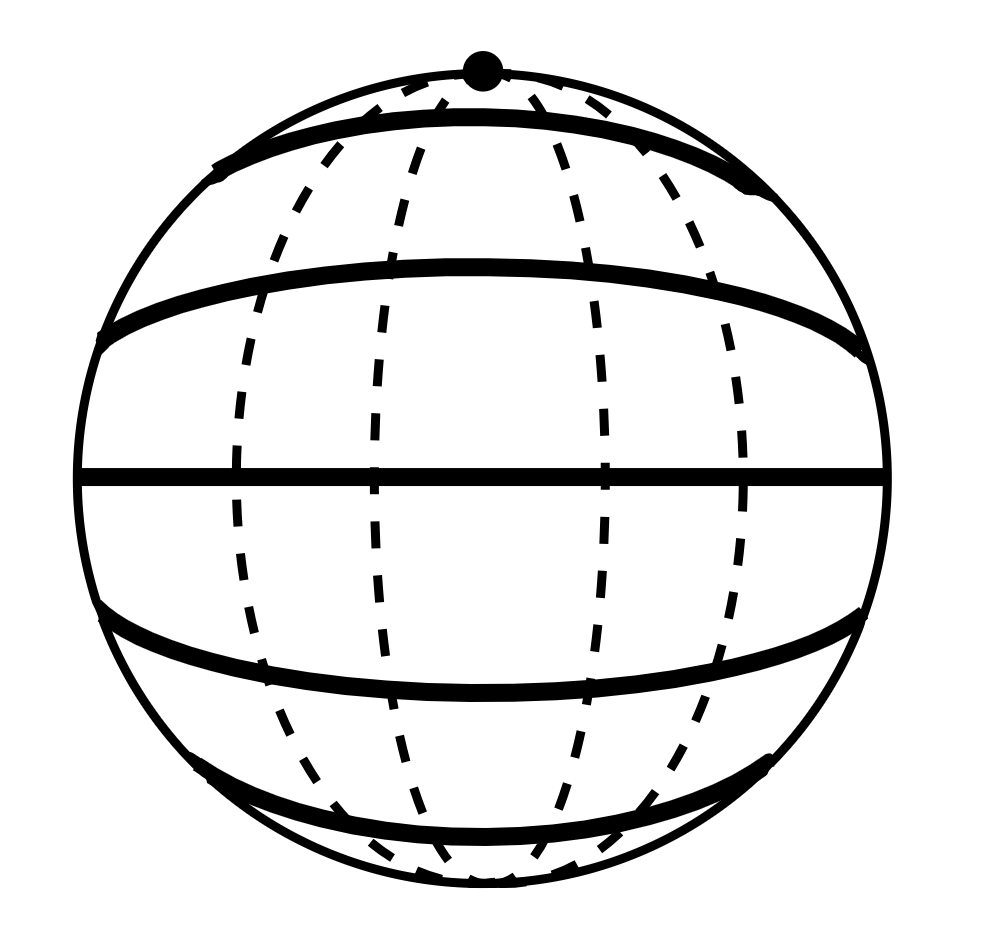}
    \caption{Unimodular coordinates $u$ (parallels) and $v$ (meridians) representation of the sphere.}
    \label{spacetime1.2}
\end{figure}

Visually, these coordinates correspond to parallels of latitude for the coordinate $u$ and meridians of longitude for $v$. See Figure \ref{spacetime1.2}.

\subsection{Static charged particle}\label{SCP}

\begin{figure}[h]
    \centering
    \includegraphics[width=0.35\linewidth]{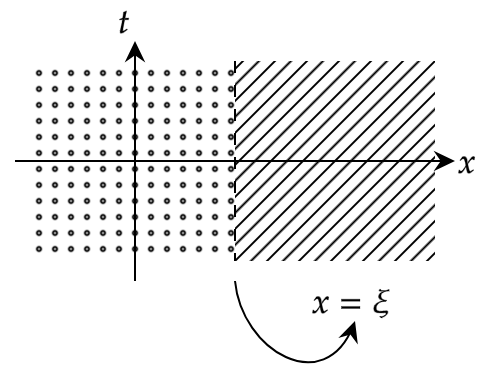}
    \caption{Electric field configuration produced by a static charged particle in flat Euclidean space at the point $x=\xi$. The positive field is represented by the hatched area and the negative field is depicted by the dotted area.}
    \label{1particle}
\end{figure}

As mentioned before, two-dimensional electrodynamics is exactly solvable. For the simple case of a static point particle with charge $q$ at a fixed point $\xi$ in the spatial coordinate, the general solution is provided in \cite{Bialynicki-Birula:1971akx} and reads:
\begin{equation}
E_q
=qH\bigl(x-\xi\bigr)+E_o\;,
\end{equation}
with $E_o$ being an integration constant while $H$ is the Heaviside function. The constant $E_o$ must be fixed by symmetry requirements, which implies the complete solution
\begin{equation}
E_q
=\frac{q}{2}H\bigl(x-\xi\bigr)-\frac{q}{2}H\bigl(\xi-x\bigr)\;,\label{campoeletrico}
\end{equation}
The representation of this field is displayed in Figure \ref{1particle}.

The discontinuity in the field means that we must perform two different mappings for different sectors and appropriately gluing the boundaries of the resulting manifolds. To the left of the particle, the curvature is negative $R=-q/2$, while to the right, the curvature is positive $R=q/2$. Therefore, the flat region to the right of the particle (a flat semi-infinite plane) can be mapped to an open semi-sphere, where the line $x=\xi$ is mapped to the equator of the semi-sphere, and the north pole is extracted. The radius of the hemisphere is then $r=2/\sqrt{q}$, according to \eqref{radius}. Thus, $\mathbb{R}^2_{x>\xi}\longmapsto S^2_{z>0}\backslash\{N\}$. 

The second half of the plane (left to the particle), however, cannot be mapped to a sphere because the curvature is negative. The corresponding manifold is actually the tractroid (pseudo-sphere), namely $pS^2$, with its cusp at infinity. Such a manifold is actually topologically equivalent to a cylinder; see \cite{Bittencourt:2025ttv,Stillwell1992,Anderson2005} and references therein. Moreover, the line $x=\xi$ is mapped to the boundary of the tractroid (the larger circumference of this manifold). In addition, this boundary is required to coincide with the boundary of the hemisphere with positive curvature. The formal map can be written as $\mathbb{R}^2_{x\leq\xi}\longmapsto pS^2\;\big|\;\partial(pS^2)=S^1_{r=2/\sqrt
q}$.

\begin{figure}[h]
    \centering
    \includegraphics[width=0.3\linewidth]{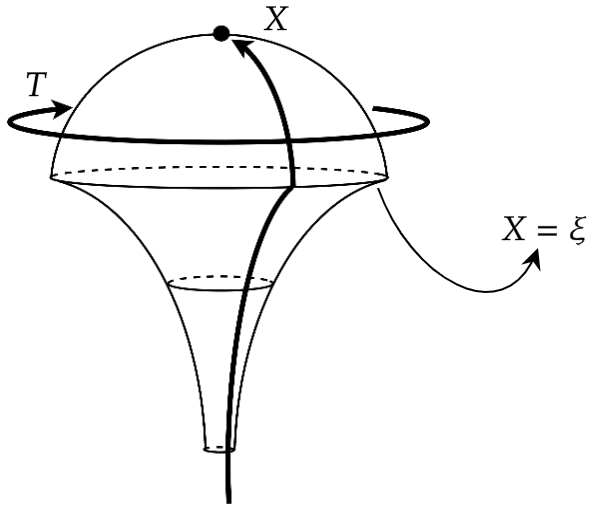}
    \caption{The ``ice cream manifold'' corresponding to the electric field of a charged point particle in flat space. The north pole is indicated by the black dot.}
    \label{spacetime2}
\end{figure}

Therefore, the resulting manifold on the gravity side will be a tractroid glued to a semi-sphere at their boundaries, \emph{i.e.}, an ``ice cream''-shaped manifold; see Figure \ref{spacetime2}. This manifold is connected, although it is not smooth.

It is clear that the line $x=\xi$ stands along the time direction (Figure \ref{1particle}). In addition, the mappings generate a compactification at the time coordinate. The world-line of the charge is thus mapped to the equator of the semi-sphere.  Thence, in the manifold in Figure \ref{spacetime2}, time is compactified and corresponds to circles parallel to the semi-sphere equator flowing clockwise (viewed from the north pole), just like the parallels of latitude. These latitude lines ``swipe'' the manifold from the north pole of the hemisphere to the cusp of the tractroid. Moreover, since the time is compactified, a period can be associated with it. The larger period is associated with the equator, and it diminishes as we move to the north pole or the cusp.

Another aspect of the mapping is the map of the particle itself. The charge $q$ must somehow be mapped in a matter configuration on the gravitational side; \emph{i.e.}, the charge density $\rho=q\delta(x-\xi)$ should correspond to a matter distribution in the corresponding gravity by means of $\rho\longmapsto\tau$. At the level of the action, the coupling between $q$ and $A$ is given by
\begin{equation}
    S_{int}\sim\int d^2x\rho\phi
\end{equation}
with $\phi$ being the electrostatic potential\footnote{No confusion with an azimuthal angle or with other fields is expected from the reader.}. Thus, the resulting matter action on the gravity side reads
\begin{equation}
    S_{int}\longmapsto S_m\sim\int d^2x\tau\omega\;,
\end{equation}
with the definition $\omega=2\omega^{01}_0$ obtained from \eqref{map1}. Because the map must relate charges to mass distributions, we can simply infer that the charge can be mapped to a massive point particle $q\longmapsto m$. Hence, the matter distribution is simply $\tau=m\delta(X)$, where we have defined the space coordinate $X$ to be orthogonal to time, and that the equator lies at $X=0$ for simplicity. As a consequence, the particle $m$ naturally describes a world-line around the equator.

It still remains to find a unimodular metric for the tractroid (for the semi-sphere, the metric is the same as in the first example). A possible metric for the tractroid is the conformally flat metric
\begin{equation}
g_{\mu\nu}=\frac{4}{q}\mathrm{sech}^2u\delta_{\mu\nu}\;,\label{gsphere2}
\end{equation}
in coordinates $u$ and $v$, for instance. 

\begin{figure}[h]
    \centering
    \includegraphics[width=0.2\linewidth]{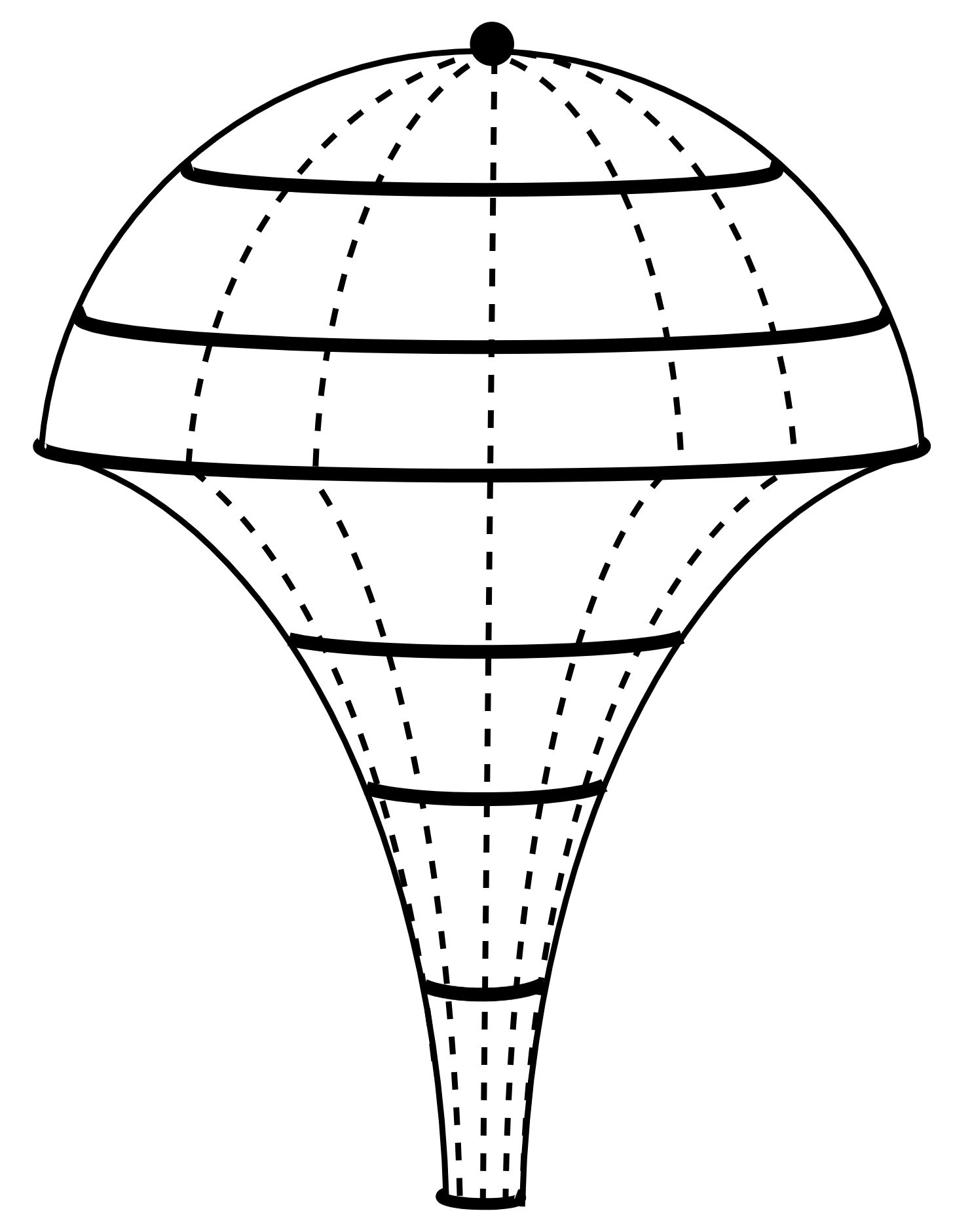}
    \caption{Unimodular coordinates $U$ (parallels) and $V$ (meridians) representation for the ``ice cream manifold''.}
    \label{spacetime2.2}
\end{figure}

The metric \eqref{gsphere2} is obviously not unimodular. The transformation to an unimodular metric is relatively simple,
\begin{eqnarray}
    U&=&\frac{2}{\sqrt{q}}\mathrm{tanh}u\;,\\
    V&=&\frac{2}{\sqrt{q}}v\;.
\end{eqnarray}
Henceforth, the new metric reads
\begin{equation}
g_{\mu\nu}\equiv\begin{pmatrix}
h(U) & 0 \\
0 & h(U)^{-1}
\end{pmatrix}\;,\label{gsphere3}
\end{equation}
with
\begin{equation}
    h(U)=1-\frac{qU^2}{4}\;.
\end{equation}
The metric \eqref{gsphere3} is clearly unimodular. Just like the unimodular coordinates $u$ and $v$ for the hemisphere, the unimodular coordinates $U$ and $V$ for the tractroid correspond to parallels and meridians, respectively. See Figure \ref{spacetime2.2}.

\subsection{Electric dipole}

\begin{figure}[h]
    \centering
    \includegraphics[width=0.4\linewidth]{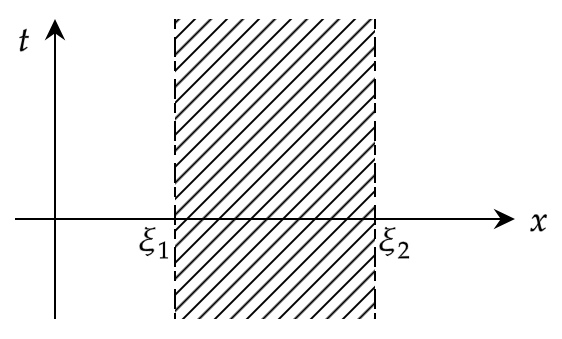}
    \caption{Electric field configuration produced by an electric dipole in flat Euclidean space at the points $x=\xi_1$ and $x=\xi_2$. The field is represented by the hatched area.}
    \label{2particles}
\end{figure}

Let us now consider the electric dipole: a positive charge $+q$ at the point $x=\xi_1$ and a negative charge $-q$ at $x=\xi_2$. Based on the previous subsection, we immediately obtain the solution
\begin{eqnarray}
    E_d
=\frac{q}{2}H\bigl(x-\xi_1\bigr)-\frac{q}{2}H\bigl(\xi_1-x\bigr)-\frac{q}{2}H\bigl(x-\xi_2\bigr)+\frac{q}{2}H\bigl(\xi_2-x\bigr)\;.
\end{eqnarray}
This configuration is schematically represented in Figure \ref{2particles}. This field correctly emulates a one-dimensional capacitor by confining the field between the charges. We have, therefore, three regions to map: $R_1=\{\mathbb{R}^2\;|\;x\in(-\infty,\xi_1]\}$, $R_2=\{\mathbb{R}^2\;|\;x\in(\xi_1,\xi_2)\}$, and $R_3=\{\mathbb{R}^2\;|\;x\in[\xi_2,\infty)\}$.

Region $R_2$ is essentially an infinite open strip of width $\xi_2-\xi_1$ where the electric field is constant. Therefore, this strip must be mapped to a sphere. This map is possible by identifying the regions $t=\pm\infty$, forming a cylinder. Then, by collapsing the two circular boundaries to a point, a sphere is obtained. However, since the boundaries at $\xi_1$ and $\xi_2$ are open, these points must be excluded and may be identified with the south and north poles of the sphere. Formally: $R_2\longmapsto S^2\backslash \{N,S\}$. 

The points at the south and north poles will naturally coincide with the positions of the particles. Time will again be compactified in terms of circles parallel to the equator, just like the previous example. Thence, the time period diminishes as one goes from the equator to a pole. At the poles, where the particles are located, the period tends to zero. Therefore, the positions of the particles can be taken as singularities where time is ill defined. 

Regions $R_1$ and $R_3$ are clearly mapped to flat manifolds because they imply vanishing curvatures. Thence, these half-planes will be mapped by an identity map to identical half-planes. One way to keep the resulting space a connected manifold  is to map each half-plane to a cone (these are topologically equivalent manifolds). Then, the right (left) closed boundary of $R_1$ ($R_3$) is collapsed to a point coinciding with the south (north) pole of the sphere. Thence, the time coordinate is compactified. Henceforth, regions $R_1$ and $R_3$ will be mapped to  two reflected cones (a cone is also locally flat).

\begin{figure}[h]
    \centering
    \includegraphics[width=0.3\linewidth]{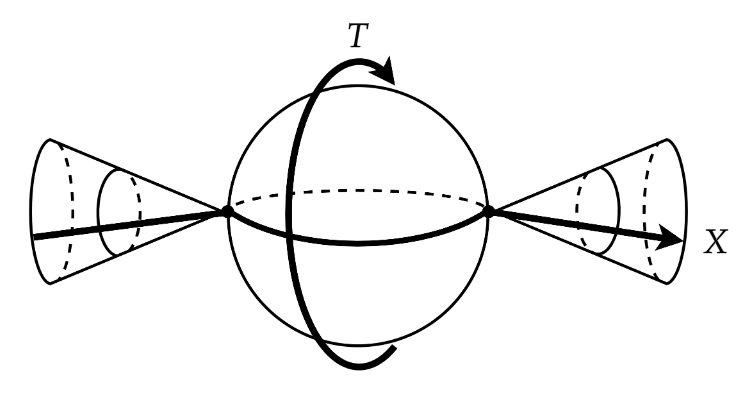}
    \caption{The ``candy manifold'': A sphere glued with two reflected cones at its poles corresponding to the one-dimensional capacitor (electric dipole). The south (north) pole, corresponding to the positive (negative) charge is to the left (right). The original lines parallel to x are mapped to meridians at the sphere and to straight lines connecting the vertex of one cone to infinity, forming a right angle to the cone cross sections. Lines parallel to time $t$ are mapped to circles parallel to the equator of the sphere, ``swiping'' the whole manifold.}
    \label{spacetime3}
\end{figure}

The corresponding manifold is displayed in Figure \ref{spacetime3}, resembling a ``wrapped candy''. Due to the fact that the mappings of all regions start with each region being identified with cylinders, it is clear that each timeline will be a parallel of latitude. The parallels ``swipe'' the whole manifold from the left cone to the right one.

The unimodular metric for the sphere is still metric \eqref{gsphere1}. For the cone, in cylindrical coordinates $(\rho,\phi)$, the metric reads
\begin{equation}
    g_{\mu\nu}\equiv\begin{pmatrix}
1 & 0 \\
0 & \rho^2\sin^2\alpha
\end{pmatrix}\;,\label{gcone1}
\end{equation}
with $\alpha$ being the included angle, \emph{i.e.}, the spherical polar angle defining the conical surface. To achieve a unimodular metric, the change of coordinates is
\begin{eqnarray}
    U&=&\frac{\sin\alpha}{2}\rho^2\;,\\
    V&=&\phi\;,
\end{eqnarray}
producing the unimodular metric of desire,
\begin{equation}
    g_{\mu\nu}\equiv\begin{pmatrix}
C(U)^{-1} & 0 \\
0 & C(U)
\end{pmatrix}\;,\label{gcone2}
\end{equation}
for
\begin{equation}
    C(U)=2U\sin\alpha\;.
\end{equation}

\begin{figure}[h]
    \centering
    \includegraphics[width=0.3\linewidth]{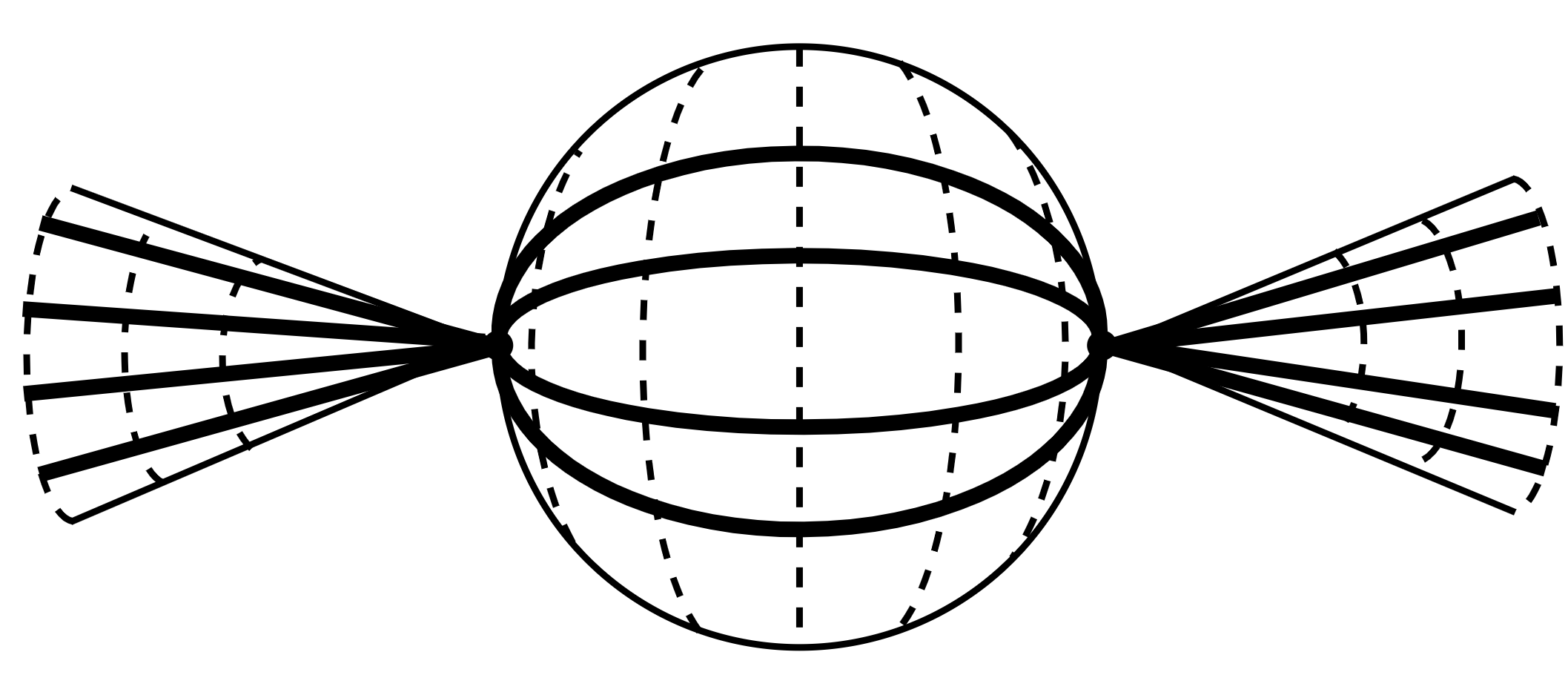}
    \caption{Unimodular representation of the ``candy manifold''. In the cone, coordinate $U$ is presented by the dashed lines (parallels) and $V$ by the solid lines (meridians). In the sphere, $u$ (parallels) is depicted by dashed lines while $v$ (meridians) is represented by the solid lines.}
    \label{spacetime3.2}
\end{figure}

Just like the unimodular coordinates $u$ and $v$ for the sphere, the unimodular coordinates $U$ and $V$ for the cone correspond to parallels and meridians, respectively. See Figure \ref{spacetime3.2}. 

As a consistency check, one should be able to deform Figure \ref{spacetime3} into Figure \ref{spacetime2} if we take away the negative charge. Therefore, we have to learn the recipe for how to make ice cream from candy. The manifolds in these figures are not topologically equivalent; extracting one charge alters the topology of the mapped manifold. Thence, the deformation of ``candy'' manifold into the ``ice cream'' manifold will require some procedures to change the topology of the manifold.

The deformation of Figure \ref{spacetime3} into Figure \ref{spacetime2} can be done as follows: i) The right cone can be contracted down to its vertex, collapsing this cone to a single point; ii) This point is then extracted. This procedure changes the topology of the manifold and is identified with the extraction of the negative charge; iii) The south pole of the sphere can be deformed into an open circle coinciding with the equator of the sphere. Hence, the sphere is deformed into a northern hemisphere with its north pole extracted. The topology of a cylinder is preserved; iii) The vertex of the left cone can be extracted, altering once more the topology of the manifold; iv) The left cone can then be deformed into a tractroid whose boundary coincides with the equator of the semi-sphere, while its boundary at infinity collapses to the cusp of the tractroid.

\subsection{Charged particle in uniform linear motion}

\begin{figure}[h]
    \centering
    \includegraphics[width=0.4\linewidth]{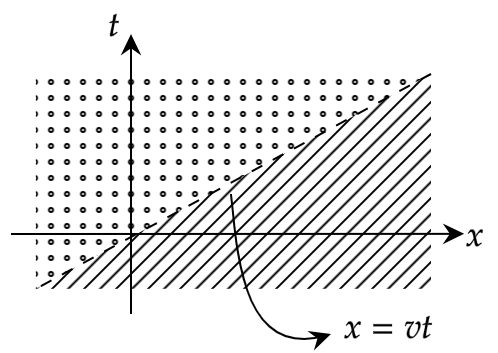}
    \caption{Electric field configuration produced by a charged particle in uniform linear motion in flat Euclidean space. The world line is set to $x=vt$. The positive field is represented by the hatched area and the negative field is depicted by the dotted area.}
    \label{mru}
\end{figure}

The last example, due to technical complications that are beyond the scope of this work, will be treated in a more qualitative manner. Let us consider a charged point particle moving in uniform linear motion to the right of the $x$ axis, with velocity $v$. The corresponding electric field reads
\begin{equation}
E_q
=\frac{q}{2}H\bigl(x-vt\bigr)-\frac{q}{2}H\bigl(vt-x\bigr)\;.\label{Emru}
\end{equation}
This field is displayed in Figure \ref{mru}. 

\begin{figure}[h]
    \centering
    \includegraphics[width=0.2\linewidth]{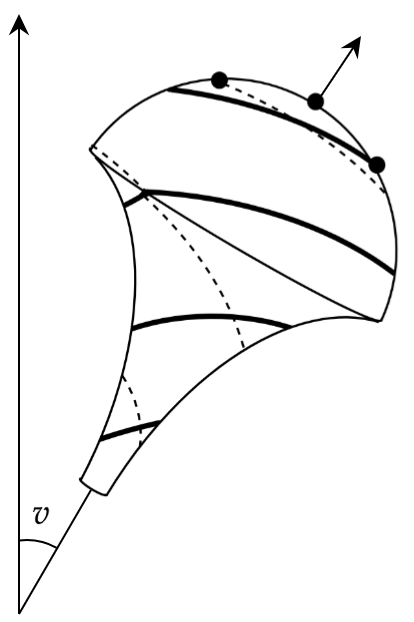}
    \caption{Rotated ``ice cream manifold'' resulting from a charged particle in uniform linear motion. The dashed lines represent the spatial coordinate $X$ and the solid lines represent the time coordinate $T$. These lines are not closed, but they curl up around the the manifold and collapse asymptotically to the north pole although, for figure be not too much noisy, we have truncated the spirals. The equator describes the particle world line on this manifold.}
    \label{spacetime4}
\end{figure}

The idea of the mapping is quite simple and can be directly obtained from the case of the particle at rest, as we will discuss in a while. The map is achieved by compactifying the world line of the particle at infinity and following the steps in Subsection \ref{SCP} (with some technical adjustments omitted here). This is essentially what we have done in the static case. However, in that case, the world line of the particle is parallel to the time axis, meaning that time is compactified in the same way the particle world line is. In the current situation, however, the world line of the particle is compactified while $t$ and $x$ follow spiral trajectories, orthogonal to each other, around the axis of the cylinder defined by the compactification of the world line. After the mapping of the planes to a hemisphere and a tractroid, the resulting manifold will be a rotation of the ``ice cream manifold'', and the coordinates following two spirals will collapse asymptotically to the north pole. The visualization is that as the particle advances in spacetime, it leaves negative curvature behind and creates positive curvature ahead. See Figure \ref{spacetime4}. 

This result is actually very intuitive because it corresponds to a Lorentz boost from the static case to the case of a moving particle. For simplicity, we take $\xi=0$ in the static field of the single particle \eqref{campoeletrico}. The only component of the electric field lies in the direction of the new frame, and thus it remains the same. In the arguments of the Heaviside functions, we replace $x\longrightarrow x-vt$, obtaining the expression \eqref{Emru}. In terms of the coordinate system, a rotation with angle $v$ is performed. This is precisely the angle enforcing the rotation of the ``ice cream manifold''.

It is quite clear from Figure \ref{spacetime4} that the position of the particle is changing over time. The mapped particle coordinates vary ($X$ and $T$), corresponding to the motion of the particle in the mapped manifold. There is a relative orientation between the particle world line and the solid lines (representing the time coordinate). In the static case, however, the particle world line lies at the equator of the ``ice cream'', and the spatial coordinate (meridians) is clearly kept constant (Figure \ref{spacetime2.2}).

\section{BRST symmetry and quantization}\label{Q}

Let us now explore the model at the quantum level. For that, we start by defining the BRST symmetry \cite{Becchi:1975nq,Tyutin:1975qk,Piguet:1995er} of the model. We will consider the non-axial starting theory, but the analysis remains the same for both cases.

\subsection{BRST symmetry and quantum electrodynamics}

The electromagnetic action \eqref{ed0} is clearly invariant under the BRST transformations
\begin{eqnarray}
    sA&=&dc\;,\nonumber\\
    sc&=&0\;,\nonumber\\
    s\Theta&=&0\;,\nonumber\\
    s\psi&=&c\psi\;,\nonumber\\
    s\bar{\psi}&=&-c\bar{\psi}\;,\label{brst1}
\end{eqnarray}
with $c$ being the Faddeev-Popov Abelian ghost field \cite{Faddeev:1967fc,Itzykson:1980rh,Piguet:1995er} and $s$ being the nilpotent BRST operator \cite{Becchi:1975nq,Tyutin:1975qk,Piguet:1995er}. The field $c$ and the operator $s$ carry ghost number $1$ and vanishing dimension. 

We also introduced the boundary term \eqref{sb0} with auxiliary fields. Being a boundary term, they might be relevant only at the boundary. Moreover, at the boundary, their field equations are algebraic. Thence, these fields are non-propagating ones. In fact, these fields can also be cast into BRST doublets, in order to reinforce that they are out of the physical spectrum of the model. In fact, these fields are compatible with the BRST transformations
\begin{eqnarray}
s\tilde{\pi}&=&\tilde{\varphi}\;,\nonumber\\
    s\tilde{\varphi}&=&0\;,\nonumber\\
    s\pi&=&\varphi\;,\nonumber\\
    s\varphi&=&0\;,\label{brst2}
\end{eqnarray}
and
\begin{eqnarray}
s\tilde{\lambda}&=&\tilde{\zeta}\;,\nonumber\\
    s\tilde{\zeta}&=&0\;,\nonumber\\
    s\lambda&=&\zeta\;,\nonumber\\
    s\zeta&=&0\;,\label{brst3}
\end{eqnarray}
We notice that the fields $\tilde{\pi}$, $\pi$, $\tilde{\lambda}$ and $\lambda$ are introduced in order to maintain the BRST doublet structure of the fields. These new fields are harmless for the model. Their quantum numbers are presented in Table \ref{table4}.

\begin{table}[ht]
\centering
\begin{tabular}{|c|c|c|c|c|}
	\hline 
Fields & $\tilde{\pi}$ & $\pi$ & $\tilde{\lambda}$ & $\lambda$ \\
	\hline 
Dimension & $0$ & $1$ & $0$ & $1$ \\ 
Rank & $(0,0,-1)$ & $(1,0,-1)$ & $(0,0,-1)$ & $(1,0,-1)$ \\
Statistics & $-$ & $+$ & $-$ & $+$ \\
\hline 
\end{tabular}
\caption{Quantum numbers of the new auxiliary fields in electrodynamics.}
\label{table4}
\end{table}

The BRST transformations \eqref{brst2} allow the auxiliary action \eqref{sb0} to be written in a BRST exact form
\begin{equation}
    S_b=s\int d\left(\tilde{\pi}\varphi-\tilde{\lambda}\zeta\right)\;.\label{sb1}
\end{equation}
Therefore, not only the fields are out of the spectrum, but also the boundary auxiliary term is irrelevant for the dynamics.

Quantization of the model can be performed defining the functional generator
\begin{equation}
    Z_{qed}=\int \left[d\Phi_{qed}\right]\exp\left(-S_{ed}-S_b\right)\;,\label{Z1}
\end{equation}
with the functional measure being the collection of the measures of all fields 
\begin{equation}
\left[d\Phi_{qed}\right]=[dA][d\Theta][d\psi][d\bar{\psi}][d\tilde{\varphi}][d\varphi][d\tilde{\zeta}][d\zeta][d\tilde{\pi}][d\pi][d\tilde{\lambda}][d\lambda]\;.\label{meas1}
\end{equation}

Of course, gauge fixing is still needed in order to have a consistent setting to perform perturbative computations. However, at this point, gauge fixing is not required.

\subsection{Mapping at quantum level}

The map of electrodynamics into gravity must now be extended to the BRST symmetries \eqref{brst1} and \eqref{brst2}. The map of the ghost field is just\footnote{BRST transformations are essentially obtained by replacing the gauge parameter $\alpha$ by the ghost field $c$. Thus, the map \eqref{map6} is naturally obtained from \eqref{map0}.}
\begin{equation}
c(x)\rightarrow\epsilon_{ab}c^{ab}(X)\;,\label{map6}
\end{equation}
one can easily check that the BRST transformations for the gravity and spinor fields read
\begin{eqnarray}
    s\omega^{ab}&=&dc^{ab}\;,\nonumber\\
    sc^{ab}&=&0\;,\nonumber\\
    se^a&=&c^a_{\phantom{a}b}e^b\;,\nonumber\\
    s\psi&=&\epsilon_{ab}c^{ab}\psi\;,\nonumber\\
    s\bar{\psi}&=&-\epsilon_{ab}c^{ab}\bar{\psi}\;,\label{brst4}
\end{eqnarray}
while for the auxiliary mapped fields are\footnote{It will be useful to infer that $s\phi^{ab}=c^a_{\phantom{a}c}\phi^{cb}-c^b_{\phantom{b}c}\phi^{ca}$.}
\begin{eqnarray}
s\xi^a&=&\phi^a+c^a_{\phantom{a}b}\xi^b\;,\nonumber\\
    s\phi^a&=&c^a_{\phantom{a}b}\phi^b\;,\nonumber\\
    s\rho^a&=&\tau^a+c^a_{\phantom{a}b}\rho^b\;,\nonumber\\
    s\tau^a&=&c^a_{\phantom{a}b}\tau^b\;.\label{brst5}
\end{eqnarray}
In this way, the fields remain as BRST doublets while the boundary term in \eqref{grav2} can be written as a BRST exact term,
\begin{equation}
    S_b=s\int d\left(\xi^a\tau_a\right)\;.\label{sb2}
\end{equation}
Again, we notice that the fields $\xi^a$ and $\rho^a$ are introduced in order to maintain the BRST doublet structure of the fields. These new fields are harmless for the model.

Applying the maps defined in Section \ref{EQUIV} to the functional generator \eqref{Z1} is not a trivial task due to non-trivial Jacobian \eqref{jac0}. Following the steps developed in \cite{Sobreiro:2021leg}\footnote{The main difference is that a correction is made here by considering $e$ inside the determinant. It is also worth mentioning that the power of $e$ in the present Jacobian is different from the one in \cite{Sobreiro:2021leg}.}, one can confirm that we can write
\begin{eqnarray}
    e^{-1/2}&=&\int[d\tilde{Z}DZ]\exp\left[-\int \left(\frac{1}{2}\tilde{Z}Z\right)\epsilon_{cd}e^ce^d\right]\;,\nonumber\\
    \mathrm{det}^{3/2}(\phi^a_{\phantom{a}b}e)&=&\int[dKd\tilde{W}dW]\exp\left[-\int\left(\tilde{W}_a\phi^{ab}W_b+\frac{1}{2}K_a\phi^{ab}K_b\right)\epsilon_{cd}e^ce^d\right]\;.\label{jac1}
\end{eqnarray}
where all new fields are 0-forms and carry 1 mass dimension. Moreover, the fields $W$ and $\bar{W}$ are ghost fields while the rest are bosonic fields.

It turns out that these local terms can also be cast into an exact BRST term. For the BRST transformations we have a BRST doublet, namely
\begin{eqnarray}
s\tilde{\chi}&=&\tilde{Z}\;,\nonumber\\
    s\tilde{Z}&=&0\;,\nonumber\\
    s\chi&=&Z\;,\nonumber\\
    sZ&=&0\;.\label{brst7}
\end{eqnarray}
and a BRST triplet, namely
\begin{eqnarray}
    s\tilde{W}^a&=&\frac{1}{2}K^a+c^a_{\phantom{a}b}\tilde{W}^b\;,\nonumber\\
    sK^a&=&-W^a+c^a_{\phantom{a}b}K^b\;,\nonumber\\
sW^a&=&c^a_{\phantom{a}b}W^b\;.\label{brst8}
\end{eqnarray}
Thus, defining
\begin{equation}
    S_{det}=\int \left(\frac{1}{2}\tilde{Z}Z+\tilde{W}_a\phi^{ab}W_b+\frac{1}{2}K_a\phi^{ab}K_b\right)\epsilon_{cd}e^ce^d\;,\label{Sdet1}
\end{equation}
we have that
\begin{equation}
    S_{det}=s\int \left(\chi Z+\tilde{W}_a\phi^{ab}K_b\right)\epsilon_{cd}e^ce^d\;,\label{Sdet2}
\end{equation}
Thence, although the fields originating from the Jacobian are not BRST doublets, $S_{det}$ is a trivial BRST co-cycle. Therefore, the Jacobian does not interfere at the physical content of the model. Once more, we notice that the fields $\tilde{\chi}$ and $\chi$ are introduced in order to maintain the BRST doublet structure of the fields. These new fields are harmless for the model. 

The quantum numbers of all new fields ($\xi$, $\rho$, $\tilde{\chi}$, $\chi$, $\tilde{Z}$, $Z$, $\tilde{W}$, $W$, $K$) are displayed in Table \ref{table5}

\begin{table}[ht]
\centering
\begin{adjustbox}{width=\textwidth}
\begin{tabular}{|c|c|c|c|c|c|c|c|c|c|c|c|}
	\hline 
Fields & $\xi$ & $\rho$ & $\tilde{\chi}$ & $\chi$ & $\tilde{Z}$ & $Z$ & $\tilde{W}$ & $W$ & $K$\\
	\hline 
Dimension & $1$ & $0$ & $1$ & $1$ & $1$ & $1$ & $1/2$ & $1/2$ & $1/2$\\ 
Rank & $(1,0,-1)$ & $(0,0,-1)$ & $(0,0,-1)$ & $(0,0,-1)$ & $(0,0,0)$ & $(0,0,0)$ & $(0,0,-1)$ & $(0,0,1)$ & $(0,0,0)$\\
Statistics & $+$ & $-$ & $-$ & $-$ & $+$ & $+$ & $-$ & $-$ & $+$\\
\hline 
\end{tabular}
\end{adjustbox}
\caption{Quantum numbers of the new auxiliary fields in gravity.}
\label{table5}
\end{table}

Thus, the final two-dimensional quantum gravity model is described by the functional generator
\begin{equation}
Z_{qg}=\int[d\Phi_{qg}]e^{-S_{qg}}\;,\label{Zqg0}
\end{equation}
with the functional measure given by
\begin{equation}
    [d\Phi_{qg}]=[de^a][d\omega^{ab}][d\psi][d\bar{\psi}][d\phi^a][d\tau^a][d\xi][d\rho][d\tilde{\chi}][d\chi][d\tilde{Z}^a][dZ^a][d\tilde{W}^a][dW^a][dK^a]\;,\label{meas2}
\end{equation}
and 
\begin{eqnarray}
S_{qg}&=&S_G+S_{det}\;,\nonumber\\
    &=&\int\frac{1}{8\pi G}\epsilon_{ab}\left(R^{ab}-\frac{\Lambda^2}{2}e^ae^b\right)+\int\bar{\psi}\left(i\gamma_ae^a\ast\nabla-\frac{m}{2}\epsilon_{ab}e^ae^b\right)\psi+\nonumber\\
    &+&s\int\left[d\left(\xi^a\tau_a\right)+ \left(\chi^aZ_a+\tilde{W}_a\phi^{ab}K_b\right)\epsilon_{cd}e^ce^d\right]\;,\label{qgrav1}
\end{eqnarray}
being the full quantum gravity action.

It is worth clarifying the scope of the present BRST analysis in relation to the axial anomaly. For nonzero fermion mass, the classical chiral symmetry is explicitly broken; nevertheless, the anomalous contribution to the axial Ward identity is still present. Schematically, the divergence of the axial current contains both the explicit mass term and the anomaly,
\begin{equation}
    \partial_\mu j^\mu=2m\bar{\psi}\gamma^3\psi+\frac{\mathrm{e}}{2\pi}\epsilon_{\mu\nu}F^{\mu\nu}\;,
\end{equation}
up to factors determined by our Euclidean conventions. In the massless Schwinger model, the anomaly generates a bosonic excitation with \(m_\gamma^2=e^2/\pi\).

The Jacobian evaluated in Appendix \ref{Ap2} is the Jacobian of the field redefinition underlying the electromagnetic-gravitational map. It must not be confused with the Fujikawa Jacobian associated with a chiral transformation of the fermionic measure. Consequently, the BRST-exactness demonstrated here does not, by itself, reproduce or eliminate the axial anomaly. Establishing its precise gravitational counterpart requires evaluating the regulated fermionic determinant in the mapped first-order background. This calculation lies beyond the scope of the present work.

We therefore conclude more narrowly that the field correspondence is compatible with the \(U(1)\)-\(SO(2)\) BRST sector, and that the auxiliary fields and the field-redefinition Jacobian do not alter its BRST cohomology. A demonstration of equivalence, including the anomalous fermionic sector, remains an open problem.

\section{Conclusions}\label{conc}

In this paper, we have enlarged the mapping established in \cite{Sobreiro:2021leg} by inserting Dirac fermions. By doing this, we could associate the two-dimensional QED action with a two-dimensional gravity action in the first-order formalism. This mapping between those theories is possible due to the well-known existence of an isomorphism connecting the respective gauge groups, namely $U(1)\longmapsto SO(2)$, as we are dealing with Euclidean spacetime. Additionally, in order to make this map bijective, we have designed it with the help of auxiliary fields to match the number of independent fields on both sides. It is important to state that a BRST analysis guarantees that these extra fields and their additional actions do not represent physical content, as we have presented. By the existence of this map, one can reinterpret two-dimensional QED as a geometrodynamical theory; it replaces Maxwell equations with a suitable configuration of spacetime deformations coupled with fermions. Therefore, in two dimensions, certain electromagnetic models described by actions involving fermions and auxiliary fields are formally related at the level of the mapped BRST sector to a two-dimensional gravitational theory with exotic fermion coupling.

As a result, we have encountered parity (P) and time reversal (T) violation through an analysis of the interaction term between fermions and spacetime in the final action. Moreover, the most interesting result of this work is that we have obtained a non-usual gauge symmetry for the fermions: it is still an $SO(2)$ symmetry, but with a $\gamma^3$ matrix which imposes that right-handed and left-handed chiral projections of the fermions transform oppositely, stating a new type of dynamics. Yet, these properties are not present on the gravitational side if we start with an axial massive Schwinger model on the electrodynamical side; thus, we can point out that our mapping establishes a connection between a P-and-T-symmetric theory and a P-and-T-violating theory. We also note that rather than dealing directly with curved spacetime, one may instead apply the map backward and study the analogous problem in a flat-spacetime theory.

In addition, we provided a few simple, explicit examples of the mapping, connecting classical electromagnetic systems to their counterparts on the gravity side. These results are displayed in Section \ref{examples}.

A particularly remarkable feature of the correspondence is the emergence of unimodular gravity on the gravitational side. This property is not imposed as an additional assumption, but follows from requiring the mapping to consistently relate gauge-invariant observables in the two theories. In this sense, the restriction to transverse diffeomorphisms is a direct consequence of the electromagnetic-gravitational correspondence rather than an independent choice in the gravitational formulation. More generally, the unimodular character of the resulting geometry can be understood as reflecting the volume element inherited from the electromagnetic theory. This observation provides an additional indication that the mapping does more than establish a formal correspondence between the two actions: it also transfers specific geometric structures and constraints from the electromagnetic description to its gravitational counterpart.

Besides carrying out a further study of the type of unusual fermions we encountered, there is still a lot to do with the mapping. First of all, checking the renormalizability of the gravitational theory, we have (it is expected to be renormalizable since the starting theory already is). Also, it would be interesting to check if the CPT theorem still holds by studying charge conjugation (C) in the gravity action. Another step to take with this map would be to apply a Wick rotation so we can deal with Minkowskian spacetime: if one were able to extend the map to $SO(1,1)$ gauge group, which is not isomorphic to $U(1)$, instead of $SO(2)$. Additionally, it is still possible to work with systems of dimension higher than two by applying dimensional reduction, as discussed in \cite{Bittencourt:2025ttv}, where three-dimensional Maxwell–Chern–Simons theory is related to a two-dimensional gravitational one by, after a suitable dimensional reduction, applying the presented mapping in the static limit. As another possible application, one could investigate this map in a more tangible system that exists in four dimensions: for instance, a graphene sheet. In this case, the starting theory would be four-dimensional QED with dimensional reduction in two spacetime coordinates. Therefore, by applying the  map, the final theory would be a two-dimensional gravity. In this way, fermions, such as electrons, living in a graphene sheet could have an alternative description: a geometrostatic one, enriching the range of condensed matter tools. Another interesting open problem concerns the confinement properties of the Schwinger model and the axial anomaly. It would be worthwhile to investigate whether the confinement features of the model admit an alternative geometrical interpretation within the gravitational formulation. Addressing this question would require extending the correspondence to nonperturbative observables, but it may ultimately provide new insights into confinement mechanisms.

It is worth mentioning that the present work consists of a formal treatment of the equivalence between electrodynamics and gravity in two dimensions, at classical and quantum levels. The present construction establishes a formal field-redefinition identity in the mapped \(U(1)\)-\(SO(2)\) sector. It should not be interpreted as an equality between fully defined, unfixed gauge-theory partition functions. An independently defined gravitational functional integral also contains a redundancy under volume-preserving diffeomorphisms, whose gauge volume has no direct counterpart in the original electromagnetic functional integral. A complete comparison therefore requires gauge fixing this residual diffeomorphism symmetry and including the corresponding ghost determinant. Such a treatment is beyond the scope of the present work. To perform actual explicit perturbative computations, gauge fixing must be considered. A traditional way to fix the gauge would be through the Faddeev-Popov method, which does not alter the path integral because it merely relies on the introduction of a ``1'' in the path integral. Additionally, gauge-invariant quantities can still be computed with no gauge fixing, at least at a formal level (see \cite{PhysRevD.74.125006} and references therein). Specifically, the gauge fixing can be done in two different ways: the first is to apply gauge fixing methods to the gravity action obtained through the mapping; the second is to apply $U(1)$ gauge fixing methods to the electrodynamical action, which induces $SO(2)$ (but not diffeomorphisms — which must be fixed as well) gauge fixing in the gravity action through the map\footnote{Notice that the gauge fixing terms are mapped onto themselves due to the structure of equation \eqref{map1}: the gauge fixing terms on the electrodynamical side are directly associated with those on the gravitational side. However, on the gravitational side, only the SO(2) part is fixed, while diffeomorphisms remain unfixed.}.

One last important point to clarify is that, since the fermionic gravitational theory presented here explicitly breaks parity symmetry, the mapping could be affected by anomalies. Therefore, a complete anomaly analysis is certainly an important direction for future investigation.

\appendix

\section{Gamma matrices in two Euclidean dimensions}\label{Ap1}

Gamma matrices are defined through the Clifford algebra\footnote{All relations are also valid for the indices $a,b,c,\ldots,h$ in tangent space of $M$.} (in our case, the Euclidean Clifford algebra),
\begin{equation}
    \{\gamma_i,\gamma_j\}=2\delta_{ij}\;,\label{gamma1}
\end{equation}
from which one can find the following Hermitian solution\footnote{This representation differs from \cite{Bertlmann:2000mw} because we are in Euclidean spacetime.},
\begin{equation}
    \gamma_0=
  \begin{pmatrix}
    0 & 1\\
    1 & 0
  \end{pmatrix}\;,\;\;\;\gamma_1=
  \begin{pmatrix}
    1 & 0\\
    0 & -1
  \end{pmatrix}\;.
\end{equation}
We also define the analogue of the 4-dimensional $\gamma_5$ matrix,
\begin{equation}
    \gamma_3=\gamma_0\gamma_1=\begin{pmatrix}
    0 & -1\\
    1 & 0
  \end{pmatrix}\;,\label{gamma2}
\end{equation}
and the generators of the spinorial representation of $\mathfrak{so}(2)$ algebra,
\begin{equation}
    \sigma_{ij}=\frac{1}{2}[\gamma_i,\gamma_j]\;.\label{gamma3}
\end{equation}
Therefore, the following relations are valid
\begin{equation}
    \sigma_{ij}=\gamma^3\epsilon_{ij}\;,\;\;\epsilon_{ij}=-\gamma^3\sigma_{ij}\;.\label{gamma4}
\end{equation}
Moreover
\begin{eqnarray}
\left\{\gamma_i,\gamma^3\right\}&=&0\;,\nonumber\\
\left[\gamma_i,\gamma^3\right]&=&2\epsilon_i^{\phantom{i}j}\gamma_j\;,\nonumber\\
\left[\gamma^3,\sigma_{ij}\right]&=&0\;.\label{gamma5}
\end{eqnarray}
Our choice of gamma matrices relates to the Pauli matrices by
\begin{equation}
\gamma^0=\sigma_1\;,\;\;\gamma^1=\sigma_3\;,\;\;\gamma^3=-i\sigma_2\;.\label{gamma6}
\end{equation}
The following properties also hold
\begin{eqnarray}
    \left(\gamma^i\right)^\dagger=\left(\gamma^i\right)^{-1}&=&\gamma^i\;,\nonumber\\
    \left(\gamma^3\right)^\dagger=\left(\gamma^3\right)^{-1}&=&-\gamma^3\;,\nonumber\\
    \left(\gamma^i\right)^2=-\left(\gamma^3\right)^2&=&1\;.
\end{eqnarray}

\section{Computation of the Jacobian}\label{Ap2}

Since there are eleven independent fields both in the 2D-electrodynamics and in the 2D-gravity constructed here, the Jacobian matrix is an $11\;\mathrm{x}\;11$ matrix. However, only the following components are non-vanishing. From \eqref{map1}:
\begin{eqnarray}
    \dfrac{\partial\Theta}{\partial e^a}&=&\mathrm{e}^2\epsilon_{ab}e^b\;, \\
    \dfrac{\partial A}{\partial\omega^{ab}}&=&\epsilon_{ab}\;.
\end{eqnarray}
From \eqref{map2}:
\begin{eqnarray}
    \dfrac{\partial \bar{\Phi}^a}{\partial\sigma^b}&=&\delta^a_b\;, \\
    \dfrac{\partial \bar{\Phi}^a}{\partial e^a}&=&\phi^a_{\phantom{a}b}\;, \\
    \dfrac{\partial \Phi^a}{\partial \phi^b_{\phantom{b}c}}&=&\delta^a_b e^c\;.
\end{eqnarray}
From \eqref{map5}, fermions are mapped onto themselves:
\begin{eqnarray}
    \dfrac{\partial \psi(x)}{\partial \psi(X)}&=&1\;,\\
\dfrac{\partial \bar{\psi}(x)}{\partial \bar{\psi}(X)}&=&1\;.
\end{eqnarray}
Noticing that some of these terms are trivial, the Jacobian matrix can be reduced to a $5\;\mathrm{x}\;5$:

\begin{equation}
J =
\begin{pmatrix}
0 & 0 & \phi & 0 &  \mathrm{e}^2e_1^1 \\
0 & 0 & 0 & \phi & -\mathrm{e}^2e_0^1 \\
\phi & 0 & 0 & 0 & -\mathrm{e}^2e_1^0\\
0 & \phi & 0 & 0 & \mathrm{e}^2e_0^0\\
e_0^1 & e_1^1 & e_0^0 & e_1^0 & 0
\end{pmatrix}\;.
\end{equation}

Finally, the Jacobian is obtained by taking the determinant of this matrix:
\begin{equation}
    \det J=-2\mathrm{e}^2{\det}^{3/2}(\phi^a_{\phantom{a}b})e\;.
\end{equation}

\section*{Acknowledgements}

The authors wish to express their gratitude to J. Zanelli, A. D. Pereira and L. E. Oxman for valuable discussions and ideas. This study was financed in part by The Coordena\c c\~ao de Aperfei\c coamento de Pessoal de N\'ivel Superior - Brasil (CAPES) - Finance Code 001.

\bibliography{BIB}
\bibliographystyle{elsarticle-num}

\end{document}